\documentclass[10pt, conference, compsocconf]{IEEEtran}
\ifCLASSOPTIONcompsoc
  \usepackage[nocompress]{cite}
\else
  \usepackage{cite}
\fi

\usepackage{graphicx}
\usepackage{verbatim}
\usepackage{xcolor}
\usepackage{url}

\ifCLASSINFOpdf
\else
\fi
\usepackage{amsmath,amssymb,amsfonts,amsthm}
\usepackage{textcomp}
\usepackage{tabularx}
\usepackage{adjustbox}
\usepackage{paralist}
\usepackage{xcolor,color,soul,graphicx}
\graphicspath{ {./figures/} }

\newcommand{\makebullet}{\noindent $\bullet$ \textbf}

\newcommand{\name}{\textsf{ETLC}}

\newcommand{\PrivBC}{\textsf{PrivBC}}
\newcommand{\PubBC}{\textsf{PubBC}}
\newcommand{\HTLC}{HTLC}
\newcommand{\VRS}{\textit{VRS}}

\newcommand{\G}{\mathsf{G}}
\newcommand{\R}{\mathsf{R}}
\newcommand{\N}{\mathsf{N}}
\newcommand{\C}{\mathsf{C}}

\newcommand{\LD}{\textsf{LD}}

\newcommand{\ELD}{\textsf{ELD}}

\newcommand{\TR}{$\R$}
\newcommand{\TN}{$\N$}
\newcommand{\TC}{$\C$}
\newcommand{\TA}{$\mathsf{A}$}

\newcommand\blfootnote[1]{%
  \begingroup
  \renewcommand\thefootnote{}\footnote{#1}%
  \addtocounter{footnote}{-1}%
  \endgroup
}

\newtheorem{claim}{Claim}[section]

\def\BibTeX{{\rm B\kern-.05em{\sc i\kern-.025em b}\kern-.08em
    T\kern-.1667em\lower.7ex\hbox{E}\kern-.125emX}}

\begin{document}

\title{Trusted Data Notifications from Private Blockchains\textsuperscript{\#}}

%\author{ % Anonymous submission
%    \IEEEauthorblockN{Paper\#87}
%}

\author{
    \IEEEauthorblockN{Dushyant Behl, Palanivel Kodeswaran, Venkatraman Ramakrishna, Sayandeep Sen, Dhinakaran Vinayagamurthy}
    \IEEEauthorblockA{IBM Research India}
}

\maketitle
\IEEEpeerreviewmaketitle
\blfootnote{\# Authors are listed in alphabetical order.}

\begin{abstract}
Private blockchain networks are used by enterprises to manage decentralized processes without trusted mediators and without exposing their assets publicly on an open network like Ethereum. Yet external parties that cannot join such networks may have a compelling need to be informed about certain data items on their shared ledgers along with certifications of data authenticity; e.g., a mortgage bank may need to know about the sale of a mortgaged property from a network managing property deeds. These parties are willing to compensate the networks in exchange for privately sharing information with proof of authenticity and authorization for external use. We have devised a novel and cryptographically secure protocol to effect a fair exchange between rational network members and information recipients using a public blockchain and atomic swap techniques. Using our protocol, any member of a private blockchain can atomically reveal private blockchain data with proofs in exchange for a monetary reward to an external party if and only if the external party is a valid recipient. The protocol preserves confidentiality of data for the recipient, and in addition, allows it to mount a challenge if the data turns out to be inauthentic. We also formally analyze the security and privacy of this protocol, which can be used in a wide array of practical scenarios.
\end{abstract}

\section{Introduction}\label{SEC:introduction}
The recent past has seen the emergence of private blockchain platforms as viable decentralized
transaction-processing systems for businesses and consortia~\cite{Fabric, Corda, quorum, sawtooth}.
Similar to public blockchain platforms like Ethereum~\cite{ethereum}, private blockchain platforms
support smart contracts running on networks of peers, updating and
recording data on a shared replicated ledger through a distributed
consensus protocol. However, network membership and access to ledger information is governed by designated network authorities, 
enabling such networks to guarantee higher privacy and accountability that are critical for enterprise scenarios.
Furthermore, a private network can use a consensus protocol suited to
the applications running on the network (e.g., PBFT~\cite{PBFT}, Raft~\cite{raft}) rather than expensive ones
like PoW~\cite{PoW}, delivering higher performance. This enables consortia to balance
performance and scalability for their specific applications by picking
and choosing the right configuration and policy parameters.
It is this customizability that makes private blockchains amenable to large scale industry adoption.

Motivated by the aforementioned benefits, a number of businesses and consortia have invested significantly to create and manage private blockchains,
running a wide range of applications including trade and supply
chains~\cite{TradeLens}, finance~\cite{wetrade, marcopolo}, regulatory compliance~\cite{ibm_kyc}, provenance~\cite{ibm_food_trust}, and real estate~\cite{BlockchainLandRegistry, BlockchainLandRegistry2}.
% to leverage the collective trust of the network instead of relying on a single trusted third party.
Network participants, typically members of a consortium, collectively control network membership and access to ledger information. As a result, private blockchain networks have turned into ``walled gardens", and the data and assets managed by them lie within silos.
But this presents a challenge to real world business processes, which typically involve far more parties that
have an interest in a private blockchain network's data than that network can accommodate
either due to scalability or privacy constraints. Clearly there is a
need for  {\bf trusted data notifications}  of private blockchain data
 to an extended set of interested parties while still respecting the privacy concerns of the private blockchain members.
%for monetary or technical reasons.
%In some instances, it may be possible to extend membership to hitherto excluded entities. But it is quite likely that a large number of scenarios exist where such expansion is undesirable or impossible for privacy or proprietary reasons. \makenote{Palani reword} Which motivates a need for 

%Specifically dictating the need of the private blockchain information for conducting business.
%unless they are willing to pay a significant amount of money to become participants, causing impediments to business.

\begin{figure}
  \centering
  \includegraphics[width=\columnwidth]{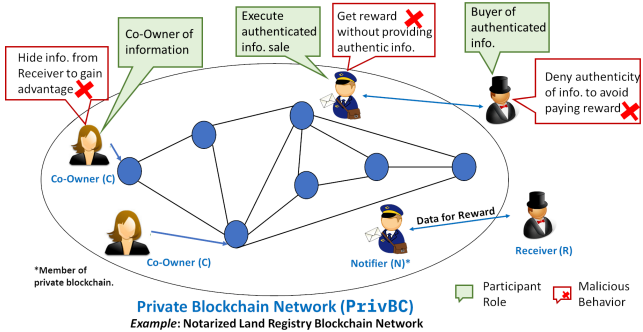}
  \caption{Private Blockchain network for real-estate ownership}
  \label{fig:motivation}
\end{figure}

To elucidate, we refer to Figure~\ref{fig:motivation} depicting a real-estate blockchain network tracking the lifecycle and ownership of property titles. The network can be constituted with membership restricted to property owners and government regulators. We note that properties registered in the network can be involved in transactions outside the scope of this network.
Land can be mortgaged or used as collateral in a business transaction, where the bank or counterparty involved is not a member of the real estate network.
A bank, for example, has no visibility into the network's ledger and does not know whether the property has changed ownership or undergone
transformation of any kind such as division amongst inheritors. Given the bank's lack of visibility into the network, a dishonest property
owner has the opportunity to defraud the bank by re-mortgaging the land to another bank.
%\makenote{PK:How is this problem solved today?} \makenote{RAMA: This problem is created by the existence of permissioned blockchains with siloes. In the pre-blockchain world, this would involve different govt agencies maintaining different sets of files, people running around, court cases}
In fact, in a blockchain ecosystem, the bank cannot rely on information from any single member of the network because ledger data is
collectively agreed upon, and a blockchain's trust emanates from this collective endorsement of data.
Property owners who have no connection with the bank may not want to expose their assets and transactions to it for privacy and business reasons.
The right approach to solve this problem is to enable the network to collectively share just the necessary data with the bank, on a need to
know basis, in a secure manner, without leaking ledger data to unintended recipients.

It is important to note that the value of the ledger data stems from it being {\it collectively agreed upon by the members of the network}.
We term this as \textit{extrinsic} value as opposed to \textit{intrinsic} value, where the data content is valuable to
external participants regardless of who supplied it. Blockchain data sans evidence of collective agreement by network members has no extrinsic
value to an external recipient, say as evidence in a court of law. 
Further, in business processes where data with extrinsic value is required, it is crucial that the data is known to have been procured through
legitimate means; i.e., the data carries information clearly indicating that the network members collectively authorize a particular non-member to use it.
This is necessary to protect the privacy and business interests of the network, and is evident in our property mortgage use case: registry network
members sharing records with unauthorized banks may face audits and penalties. This condition becomes important in our solution design as described in later sections.

Note that usage of trusted data notifications is well-entrenched and legally accepted in modern economic and commercial processes. Examples include getting electronic signatures endorsed by respective agencies~\cite{ec_esignatures},
%account statements of an individual endorsed by the respective banks,
creditworthiness statements endorsed by ratings agencies~\cite{fitch_credit_rating}, warrants of physical presence of goods endorsed by warehouses~\cite{intracen_warehouse_recipt}, and documents of ownership of real estate (or other assets) endorsed by respective agencies~\cite{eu_land_registry}.

Unfortunately, no methods currently exist for private networks to supply trusted notifications of extrinsically valuable data to outsiders.
{\it Members of private blockchains} as a {\it group} can themselves act as trusted sources of data and provide notary services to external members in specific scenarios~\cite{blockchain_notary, blockchain_credit}. As an added incentive, supplying data in exchange for money may also enable a consortium's members to monetize and recoup their investments in maintaining a blockchain network.

Therefore, we set out to create {\it trusted} mechanisms to \textit{enable a private blockchain network to share private ledger data authorized for a non-member by the network securely and confidentially in exchange for a monetary reward.} In our model, protecting the privacy and business interests of the network is crucial: the consortia members should collectively endorse not only the veracity of data but also the set of external members with whom that data can be shared in an authorized manner. 
%Examples{\bf XXX cites from each of the examples below} include \textcolor{red}{KYC validations, background checks for criminal behavior,} getting account statements of an individual endorsed by the respective banks, creditworthiness statements endorsed by ratings agencies, \textcolor{red}{warrants} of physical presence of goods endorsed by warehouses, documents of ownership of real estate (or other assets) endorsed by respective agencies.
The solution we will propose relies on existing mechanisms to generate collective endorsements, typically using digital signatures~\cite{Fabric}.

% Read this paragraph from Rama's version and check if something got missed from this.

%We now define and scope our problem: delivering {\bf trusted data notifications} from private networks to non-members ("trusted data" refers to data accompanied by certifications and authorizations, and possessing extrinsic value to its seeker).
%The use of such notifications is well entrenched and legally accepted in modern economic, commercial processes. Examples{\bf XXX cites from each of the examples below} include \textcolor{red}{KYC validations, background checks for criminal behavior,} getting account statements of an individual endorsed by the respective banks, creditworthiness statements endorsed by ratings agencies, \textcolor{red}{warrants} of physical presence of goods endorsed by warehouses, documents of ownership of real estate (or other assets) endorsed by respective agencies.
%Our motivation to tackle this problem has not been addressed in literature but will soon become a critical need in the blockchain ecosystem.
%We can immediately identify another real-world need: that private networks may have no incentive to notify non-members in the absence of regulation. But a monetary incentive, which would make {\it members of private blockchains} act as a paid {\it group} notary service has a very high likelihood of serving the needs we have listed. Payments in exchange for data would also allow private blockchain administrators to cover network setup and maintenance costs.

Our problem is essentially a variant of {\it Fair
  Exchange}~\cite{asokan_fair_exchange} between two untrusting
parties: private blockchain network members (as a collective) and an external data recipient.
Each party faces the risk of being cheated by its counterparty
unless precautions are taken. For example, a naive solution in our property scenario would involve
the interested bank periodically querying peers (owned by members) of the private network
(at least a majority) to get updates on properties of interest. But this solution is neither scalable (because of polling overhead) nor lucrative (lack of incentives for members to share data). An alternate solution, where the bank is notified by any member of the private network in return for a reward, is no better as the member may supply fake data or the bank may renege on making a payment.

Two-party fair exchange is known to be infeasible in the absence of a trusted third party~\cite{FairExchangeInfeasibility}. The solution that we will present uses a public blockchain with programmable money constructs as a trusted arbitrator to guarantee fairness for both the external recipient (like a bank) and the private network member group. In our solution model, a transaction (data in exchange for money) is executed by a \textit{Notifier}, a role that can be assumed by any member of the private network to supply data and earn a pre-determined reward (for itself and members of the private blockchain) if the data is authentic (see Figure~\ref{fig:motivation}). At the heart of our solution lies a pair of interlocking contracts to ensure that the receiver gets authentic information (containing private blockchain network endorsements) if and only if all private network members get their reward. Our protocol, while inspired by Hashed Time Lock Contracts(\HTLC)~\cite{HTLC}, ensures authenticity with confidentiality by leveraging zero knowledge and encryption mechanisms.
%\makenote{PK: Intro should give some more of solution outline. Not just one line}

%\newline
\noindent{\bf Key contributions:}
In this paper, we make the following contributions
1) We motivate and present the design of our solution for trusted data
notification, {\it Evidence Time Locked Contracts (\name)} which uses zero knowledge proofs and standard cryptographic techniques
to guarantee fairness in the exchange of private blockchain data with external authorized
recipients. 2) We formally define the claims made by our
solution and present a security analysis of our protocol. In
particular, our solution guarantees fair exchange between the private network members' group
and the receiver, and confidentiality of blockchain data 3) Our solution
further guarantees that the network's privacy policies are always
enforced and even a rogue notifier cannot exfiltrate valid notifications to unauthorized
recipients.
%\makenote{TODO for Dhinakaran:Reword contributions}
%security properties for \name. \makenote{XXX security claims here}. \makenote{XXX Any other key contributions ?}.

We organize the rest of the paper as follows. First, we discuss our problem setting and list the building blocks for our solution in Section~\ref{SEC:setting_and_prelims}.
We describe our protocol \name\ in detail in Section~\ref{SEC:solution}. We formally analyze the correctness, security, and privacy features of the protocol in Section~\ref{SEC:analysis}, survey related work in Section~\ref{SEC:related}, and conclude with our ideas for future research in Section~\ref{SEC:conclusion}.

\section{Problem Setting and Preliminaries}\label{SEC:setting_and_prelims}
In this section, we model the problem and describe the building blocks of our proposed solution.

\subsection{Problem Setting}\label{SEC:building_blocks}
% \begin{figure}[t]
%     \centering
%     \includegraphics[width=0.9\columnwidth]{permissioned_bc.eps}
%     \caption{Model of a private blockchain network}
%     \label{FIG:permissioned_bc}
% \end{figure}
A private blockchain network, denoted by \PrivBC, is collectively governed by a group of entities (sometimes called \textit{consortium}) that need to cooperate for business reasons but do not completely trust each other.
Though the technologies they are built on vary widely in structure and function~\cite{Fabric, Corda, quorum, sawtooth}, we can model them commonly as
$<${\textsf{Peers},\textsf{SC},$\textsf{C}$,$\textsf{L}$,$\textsf{M}$}$>$, where:
\begin{compactitem}
\item \textsf{Peers} denotes the peers belonging to different business entities (sometimes called \textit{organizations})
\item \textsf{SC} denotes the smart contracts running business logic
\item $\textsf{C}$ denotes the consensus protocol to achieve finality and ordering of transactions
\item $\textsf{L}$ denotes the ledgers maintaining chains of blocks and state snapshots
\item $\textsf{M}$ denotes membership policy, governing entry and access to \PrivBC
\end{compactitem}
A transaction is typically submitted by an application client, pre-authorized using $\textsf{M}$, to the network.
Signatures (endorsements) are collected from a subset of \textsf{Peers} (the list is determined by consensus policy) running appropriate \textsf{SC} functions.
Subsequently, the transaction is validated using pre-configured consensus policies before it is committed independently by each peer to its replica of a ledger.
Such a commitment consists of a given ledger data \LD\ item with an updated value, and this is stored in the ledger along with peers' endorsements.

% 2) The client collecting endorsements from other peers with the goal of satisfying the endorsement policy of the smart contract. Endorsement entails the peers simulating the smart contract execution and signing the outputs with their private keys.
% 3) Finally the client submitting the endorsed proposal to the blockchain consensus protocol for inclusion in the blockchain.
% 4) The consensus protocol performs a number of validation checks to ensure the endorsed transaction is well formed before committing to the blockchain ledger.
The consensus protocol $\textsf{C}$ ensures that data committed to the blockchain
represents the consortium's collective will, the proof of which lies in the endorsements associated with ledger data (\LD).
%Such authenticated data can play an important role in resolving disputes among organizations that are outside the consortium (or do not satisfy the membership policy \textit{M}) or involved in inter-related business processes.
%\makenote{Have a strong story in intro with banking, mortgage that extends beyond consortium}
%\makenote{XXX: Why do we need this much detail of fabric? can't we abstract out the endorsement by saying signing and be done}
To an external entity, \LD\ coupled with endorsements may possess extrinsic value (see Section~\ref{SEC:introduction}),
however, the membership policy $\textsf{M}$ ensures that \PrivBC\ ledger
%data is accessible only to the consortium's members, thereby severely reducing the value of this authenticated data to external members who might be interested.
data is accessible only to the consortium's members.
%On the other hand, including every possible interested external member in the blockchain not only violates the privacy/scalability constraints of the network, but also adds additional overhead of updating and revoking access to external members after their need for access expires. 

One approach to make \LD\ available outside the network is for \PrivBC\ to expose a \textit{Verifiable Read Service (\VRS)}~\cite{middleware_interop} API that allows external members to query for a given \LD.
In response, \PrivBC\ will return \LD\ along with endorsements and other metadata if the requestor passes an access control check enforced collectively by \textsf{Peers}.
%On receiving a request, the peer verifies an access control policy and obtains consent from the rest of the peers
%in the network. We reiterate that \LD\ is valid outside the network
%for a Receiver \TR\ only if it contains endorsements(signatures) for
%both its validity as well as authorizations for \TR. The
%authorizations ensure that the consortium members as a collective
%still retain control on access to private blockchain data to external
%members. \makenote{PK:reword} The \VRS\ response consists of \PrivBC\
%ledger data including the validity and authorization endorsements
%signed by multiple peers based on the network policy.
However, the use of a \VRS\ raises two challenges: (i) the peers lack incentive to offer such a service freely to external members, and
(ii) the pull model requires the external member to continuously poll for updates from \PrivBC.
In our solution \name, we address (i) using an incentive mechanism for network members, and
(ii) using a notification based push model.
In this push based model, any member of \PrivBC\ may assume the role of notifier \TN, sharing updates to \LD\ with a designated external receiver \TR.
In return, \TR\ pays a reward to the members of \PrivBC\ after verifying the authenticity of the data.
This now reduces to a fair exchange
problem~\cite{asokan_fair_exchange} between the \TN\ and \TR\,
requiring the use of a trusted third party arbitrator to guarantee
fairness~\cite{FairExchangeInfeasibility}. In \name, we leverage a public blockchain network
(accessible to \TR\ and members of \PrivBC) as a trusted arbitrator, exploiting its
programmable money constructs to guarantee fairness of the ledger data-reward exchange.

The core exchange mechanism we will use in \name\ is inspired by \textit{Hashed Time Lock Contracts} (\HTLC)~\cite{HTLC}, a mechanism that allows two parties on two different public blockchains to swap assets with each other. % using an agreed-upon exchange value.
\begin{comment}
In an \HTLC\ instance, one of the parties deploys a contract on one network promising to transfer an asset to the other party when a secret preimage $s$, known only to it and corresponding to an encoded hash, is supplied. The other party deploys a similar contract encoded with an identical hash on the other network, promising to transfer another asset to the first party.
In an \HTLC\ instance, parties pledge assets to each other through contracts deployed on the two networks,
each encoded with the hash $H(s)$ of the same preimage $x$, known only to one of the parties a priori.
($H(\cdot)$ is a standard hash function like SHA-256.)
Upon production of a secret $s$ that matches $H(s)$, transfer of assets happen in both directions without either party getting an opportunity to renege on its pledge (hence {\it hash lock}).
\end{comment}
In an \HTLC\ instance, one of the parties deploys a contract on one network promising to transfer an asset to the other party when a secret preimage $x$, known only to it and corresponding to a hash $H(s)$ encoded in the contract, is supplied ($H(\cdot)$ is a standard hash function like SHA-256). The other party deploys a similar contract encoded with the identical hash on the other network, promising to transfer another asset to the first party.
Upon production of a secret $s$, the hash of which matches $x$, transfer of assets happen in both directions without either party getting an opportunity to renege on its pledge (hence {\it hash lock}).
These contracts are also {\it time locked} i.e. if either party fails to submit a transaction with the right preimage within a given time $t$, the escrowed asset is returned to its owner.

\subsection{Trusted Data Notification Model}\label{SEC:model}
We now formally describe the \PrivBC\ trusted data notification problem and assumptions of our threat model.
\newline
We consider the problem of enabling an external receiver \TR\ (that is not a member of \PrivBC) to be notified of a blockchain event $\mathsf{e}$ of
interest in return for paying a reward.
In our current model $\mathsf{e}$ represents the creation, updation, or deletion of a specific ledger data \LD\ item that \TR\ is interested in.
We define \textit{Notifiers} $\{\N\}$ (consisting of \PrivBC\ members) to be any set of
entities that notify $\mathsf{e}$ to \TR\ in exchange for a reward $\mathsf{a}$.
%\PubBC\ is a public blockchain such that \TR,\TN\ and all \textsf{Peers} of \PrivBC\ own accounts and can transfer money (native assets, like cryptocurrency) with each other on the network.
\PubBC\ is a public blockchain such that \TR, \TN\ and all members of \PrivBC\ (including \textsf{Peers} and clients) own accounts and can transfer money (native assets, like cryptocurrency) to others on the network.
%We use a public blockchain \PubBC\ %(s.t. $\N,\R\subset\PubBC$)
%as a mediating platform for this exchange between \TN\ \& \TR.
Importantly, we assume that $\mathsf{e}$ holds {\it extrinsic} value for \TR\ and hence carries utility only if it can be proven
to (i) emanate from \PrivBC\ and (ii) authorized by \PrivBC\ for \TR.
%\makenote{RAMA: We talked about this in Sec 1 already, why do we need to repeat it here?}
\begin{comment}
  Consider a permissioned blockchain network \PrivBC, $\G\subset\PrivBC$ is a client which creates and updates a particular asset(henceforth referred to as \textit{data}) on the \PrivBC. \TR\ is an external party which is interested in creation/updation of this private blockchain {\it data} and is ready to pay a reward in exchange. In this setting, we define $\{\N\}\subset\PrivBC$ to be a set of designated entities called notifiers(\TN) which assume the role of executing the private blockchain data sale to \TR.
A naive solution that guarantees R will be notified of
\emph{all} events $e$ of interest involves making R a member of the
blockchain. However, this solution is not feasible due to  private
blockchain privacy and scalability constraints.  Typically, a private blockchain is typically run by a  small consortium of entities whose  mutual business interests benefit participation in the network. Such networks typically restrict participation to only those entities that provide mutual benefit to other members of the network. Receiver R, on the other hand being a consumer of data does not provide benefits to other members and hence will not satisfy membership policy $P$ of the network. \textbf{XXX: Model text should talk about membership policy P}. Furthermore, private blockchains run pbft variants of consensus protocols that do not scale for large number of entities, thereby preventing R from becoming a member of the network.
\end{comment}

\subsubsection{Threat Model}\label{SEC:threat_model}
We make the following assumptions:

\makebullet{\TR\ is rational:} Receiver \TR\ has a genuine interest (economic or otherwise)
in receiving % subscribing to
event $\mathsf{e}$ and has no incentive to broadcast private blockchain data to others.
\newline
\makebullet{\TN\ is rational}: Notifier \TN\ is incentivized to notify \TR\ about $\mathsf{e}$ if reward is guaranteed in return.
%provided \TN\ can be guaranteed of rewards for successful delivery.% \makenote{RAMA: Can we add here that \TN\ has no disincentive to do this, as everyone staying silent means no reward, whereas anyone speaking up means everybody gets a reward?}
\newline
\makebullet{\PrivBC\ is trustworthy}: The private blockchain is viewed as a collection of peers trusted to finalize transactions and
record data on the ledger using an advertised % endorsement policy and \makenote{XXX: endorsement policy is not explained so remove from here}
consensus protocol, even if some peers may act dishonestly.
We assume that the members of \PrivBC\ will not collectively collude
to cheat \TR\ by supplying fake data or endorsements.
We also assume the membership service of the network is trustworthy and issues valid certificates to the \PrivBC\ members.
\newline
\makebullet{\PubBC\ is trustworthy}: The public blockchain's consensus protocol is trusted to accurately execute smart contracts and
record transactions. % in the blockchain.
We also assume that \PrivBC\ members cannot collude and corrupt the consistency of \PubBC.

%\makenote{Anything about consistency models and finality? [RAMA: No need IMO]}
\subsubsection{Goals}\label{SEC:goals}
Our protocol must ensure the following:

\makebullet{Fair Exchange }:
Each member of \PrivBC\ (including \TN) must obtain a reward in \PubBC\ for every valid $\mathsf{e}$ that is successfully delivered to \TR. This protects \TN\ from a
malicious \TR\ that receives an endorsed \LD\ but later denies
payment to \TN\ by claiming non-delivery. Similarly, \TR\ is protected against a malicious \TN\
that delivers an $\mathsf{e}$ that either (i) does not represent ledger data or (ii) has no associated endorsements
from peers proving validity or authorization.
%In this case, a malicious \TN\ notifies \TR\ of an event $\hat{e}$ that is not committed to ledger
%of \PrivBC.
\newline
%\makebullet{Reward Guarantee}: In this case, a malicious \TR\ does not pay the notifier \TN\ on notification of valid event $e$.
\makebullet{Confidentiality}: A notification can be
delivered only to an \TR\ that is authorized by \PrivBC\ to
receive notifications. This prevents unintended recipients from
receiving an \LD\ carrying endorsements proving both its validity and the network's authorization for \TR.
Since we are using a public blockchain as an arbitrator, this is a salient concern.

%\textcolor{red}{Additionally, we also satisfy the properties of {\it information assurance}
%and {\it notification receipt undeniability} as described in section~\ref{SEC:analysis}.}
%\makenote{PK: Do we explicitly specify we do not protect against data only being committed to ledger --ie. lack of authorizations}
%\makenote{RAMA: We can't make any guarantees of timeliness. We can add a note in the rationality and honesty assumptions saying that since every notifier is rational and they are not all colluding to deny timely updates, an update will get delivered in a timely manner by implication.}
%%%%%\makenote{XXX: Why not mention all 4 attacks which the analysis talks about here?}

\section{Solution: Trusted Data Notification Protocol}\label{SEC:solution}
In this section, we will describe our protocol \name\ for the
notification of valid ledger data from a private blockchain (\PrivBC)
to an intended external recipient (\TR) in exchange for a reward.

\subsection{Smart Contracts}
%\makebullet{Smart Contracts:}
\name\ utilizes a set of smart contracts deployed on \PrivBC\ and the
\PubBC. These contracts embed logic to record and verify various
artifacts of hashes, ciphertext, keys, signatures etc. to enable the
guarantees provided by our solution. For instance, some of these
contracts utilize {\it hash and time locks} as discussed in
Section~\ref{SEC:building_blocks}, supported by public blockchains
like Ethereum~\cite{ethereum} to guarantee liveness of the protocol.
The list of contracts and their functions are listed in
Table~\ref{tab:contracts}. We will now describe how the contracts are used in our protocol construction.

\begin{table}[th]
  \caption{System Contracts}
  \label{tab:contracts}
  \begin{adjustbox}{width=0.48\textwidth}
  \centering
  \begin{tabular}{|m{1.4cm}|m{1cm}|m{4cm}|m{2.6cm}|}
  \hline
  \textbf{Name} & \textbf{Network} & \textbf{Description} & \textbf{Operations} \\
  \hline
  SC-ACL & PrivBC & Maintains access control list: \textit{subjects} are external recipients (identified by their public keys) and \textit{objects} are ledger data elements. Encrypts updated ledger data for external notifications. & \textit{PermitAccess}, \textit{RevokeAccess}, \textit{RecordEncDataHashProof} \\
  \hline
  SC-Reward & PubBC & Guarantees rewards to notifiers after due verification of the recipient having received authentic data, utilizing HTLC techniques. & \textit{RecordPubKey}, \textit{RecordCipherTextHashProof}, \textit{RecordSignature}, \textit{VerifySignature}, \textit{Validate} \\
  \hline
  SC-R-Sign & PubBC & Guarantees amount in exchange for authentic signature submitted before timeout. & \textit{RecordSignature} \\
  \hline
  SC-N-Key & PubBC & Guarantees amount (identical to SC-R-Sign) in exchange for valid decryption key (symmetric) submitted before timeout. & \textit{RecordKey} \\
  \hline
  \end{tabular}
  \end{adjustbox}
\end{table}

%\subsection{End-to-End Protocol Design}\label{SEC:solution-e2e-protocol}
\subsection{\name\ Protocol}\label{SEC:solution-e2e-protocol}
\begin{figure}
  \centering
  \includegraphics[width=\columnwidth]{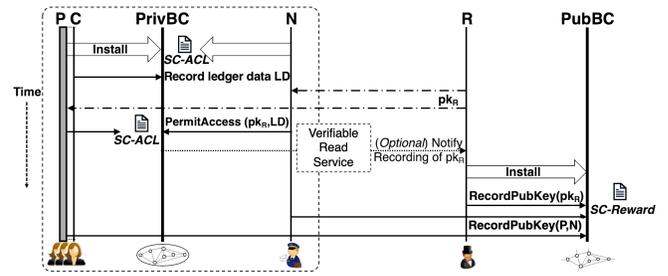}
  \caption{End-to-End Protocol: Bootstrap Stage}
  \label{fig:protocol-bootstrap}
\end{figure}

\begin{figure}
  \centering
  \includegraphics[width=\columnwidth]{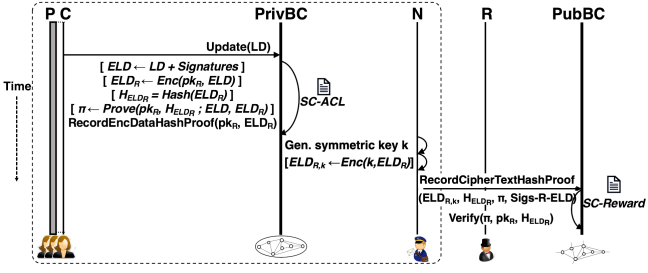}
  \caption{End-to-End Protocol: Generation Stage}
  \label{fig:protocol-generation}
\end{figure}

In \name, every unique external recipient of a private ledger data element goes through four sequential stages, as illustrated in Figures~\ref{fig:protocol-bootstrap}--\ref{fig:protocol-verification-reward}.

Note that in each stage diagram, we use \textsf{P} to denote all members of \PrivBC, excluding \TN; \TC\ (Co-Owner) is any member in \textsf{P} that creates or updates ledger data (\LD) of interest to \TR. Any member of \PrivBC\ can assume the role of \TN\ in any given instance without requiring changes to the protocol.
For rewarding purposes, we will assume that every member in \textsf{P} has an identity and crypto-currency account in \PubBC\ for conducting transactions.
%(In future extensions, we may consider grouping these members by the organizations units they represent within \PrivBC.)
%Let us now examine the processes in each stage in more detail.
We now describe each process in more detail.

%\makebullet{Bootstrap Stage:}
\subsubsection{Bootstrap Stage}
The \textit{Bootstrap stage} need only be done once for a given
$\langle$recipient, \LD$\rangle$ pair. In Figure~\ref{fig:protocol-bootstrap} first, the
\textit{SC-ACL} contract must be installed in \PrivBC\ through the
network's consensus process. This needs to be done just once in the lifecycle of the network.
This contract supports the registration of external recipients' public
keys (or certificates) for identification and cryptographic purposes.
We assume that recipient \TR\ can communicate its public key
($pk_{\R}$) separately to every member of the network as indicated in
Figure~\ref{fig:protocol-bootstrap}, and that the key is committed
through consensus to ledger using \textit{SC-ACL}'s
\textit{PermitAccess} operation. Optionally, \TR\ may submit $pk_{\R}$ within a certificate signed by authorities that can be validated in this operation. As the figure also indicates, we assume the data item that \TR\ has been given access to (\LD) is already recorded on the ledger.

\TR\ can confirm through a \VRS\ (see
Section~\ref{SEC:building_blocks}) that (i) $pk_{\R}$ has been associated
with \LD\ on \PrivBC's ledger, and (ii) that \PrivBC
has authorized \TR to receive any future updates made to \LD.
%and its future received notifications have extrinsic value. 
%is highly desirable for efficient operation, though not mandatory; \TR\ does not face any risk by carrying out the subsequent steps, as we will prove later in Section~\ref{SEC:analysis}. The action now turns to \PubBC. \TR\ is now ready to install the \textit{SC-Reward} contract in anticipation of data update notifications. $pk_{\R}$ is recorded using the \textit{RecordKey} operation as are the public keys of \PrivBC\ members (all of \textsf{P}, including \TN), which may be required for authentications later in the protocol, on \PubBC's ledger using the \textit{SC-Reward} contract.

%\makebullet{Generation Stage:}
\subsubsection{Generation Stage}
The goal of the generation step is for
\TN\ to commit to the contents of the notification while also proving
(using a zero knowledge proof) that the notification contains valid
updates from the ledger.
When \LD\ is updated on \PrivBC's ledger by any \TC, a call to \textit{SC-ACL} is triggered using \PrivBC's transaction submission mechanism. At this point, \LD\ already has signatures endorsing it on the chain (ledger), so this call can package \LD\ with the signatures into \ELD\ (or endorsed \LD). The following are produced and recorded on the ledger using the \textit{RecordEncDataHashProof} operation, as illustrated in Figure~\ref{fig:protocol-generation}:
\begin{compactitem}
\item $\ELD_{\R}$: a deterministic public-key encryption of \ELD\ using $pk_{\R}$ ensuring only \TR\ can decrypt the notification
\item H$_{\ELD_{\R}}$: a hash of $\ELD_{\R}$ as a commitment\footnote{a randomized commitment scheme is used, with the randomness/opening to the commitment recorded on \PrivBC{} and later recorded on \PubBC{} during the KeyTransfer stage.} % that acts as a commitment
\item $\pi_{\ELD_{\R}}$: a zero-knowledge proof that can be used to verify, without knowing $\ELD_{\R}$, that $\ELD_{\R}$ was produced by encrypting some plaintext under $pk_{\R}$.
\end{compactitem}
We will use the Encrypt-with-Hash construction in Bellare et al. \cite{BBO07} as our deterministic encryption scheme with the ElGamal encryption scheme \cite{Elgamal85} as the underlying randomized public-key encryption scheme. This deterministic encryption provides PRIV-CCA security \cite{BBO07} and it's good enough to provide confidentiality to \ELD\ since the underlying unforgeable signatures are `unpredictable'. For this encryption, the zero-knowledge proof $\pi_{\ELD_\R}$ is generated using a pair of discrete-log exponent checks followed by a hash check. 

\TN\ now privately generates a symmetric key $k$ (we use AES-CTR\footnote{IND-CPA security provided by AES-CTR is sufficient as we show in section \ref{SEC:analysis}}) and further encrypts $\ELD_{\R}$ into $\ELD_{\R,k}$. This second layer of encryption is to prevent $\R$ from learning $\ELD_\R$ (and hence $\ELD$) before the next step. On \PubBC, \TN\ records the following through the \textit{RecordCipherTextHashProof} operation on the \textit{SC-Reward} contract:
\begin{compactitem}
\item Doubly encrypted data ($\ELD_{\R,k}$)
\item Hash H$_{\ELD_{\R}}$ of the singly encrypted $\ELD_{\R}$, which is recorded on \PrivBC's ledger after consensus among its members
\item Zero-knowledge proof $\pi_{\ELD_{\R}}$
\item Signatures validating \PrivBC's collective endorsement of \TR's authorization to access \LD: {\it Sigs-R-ELD}; generated on the private chain upon successful conclusion of the \textit{RecordEncDataHashProof} operation
\end{compactitem}
The proof $\pi_{\ELD_{\R}}$ is now verified using $pk_{\R}$, which was recorded on the ledger in the bootstrap stage, and H$_{\ELD_{\R}}$. If the verification succeeds, i.e., the proof and the hash are authentic, it implies that the hash was produced from data which itself was generated by encrypting a different data element using $pk_{\R}$. (We will see why this is useful when the preimage of the hash is revealed in the next stage.) If the verification fails, the protocol terminates at this stage.

%\makenote{Move the multiple notifiers to discussion}
%As an aside, we remarked earlier that there is no restriction on any member of \PrivBC\ acting as \TN, and it is possible that many may choose to do so upon an update to \LD. So \textit{RecordEncDataHashProof} may be invoked by multiple \TN s on \textit{SC-Reward}, which must pick one for subsequent operation and reject the others. Managing contention is a challenging issue, though one familiar to distributed system practitioners; hence, we will not go into details in this paper apart from suggesting that an additional version attribute (to reject older updates to \LD) and a first-come-first-serve selection policy for an \TN\ may serve the purpose. In subsequent stages, we will assume a single \TN.

The remaining steps of the protocol occur on \PubBC\ and do not involve \PrivBC.

\begin{figure}
  \centering
  \includegraphics[width=\columnwidth]{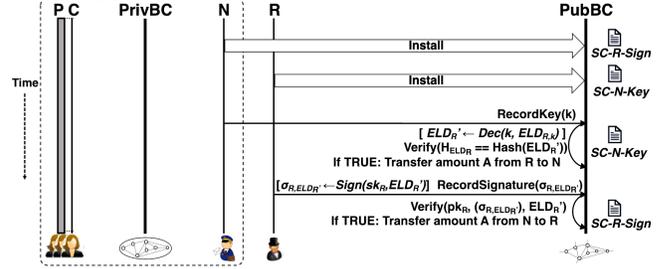}
  \caption{End-to-End Protocol: Key Transfer Stage}
  \label{fig:protocol-key-transfer}
\end{figure}
\subsubsection{Key Transfer Stage}
%\makebullet{Key Transfer Stage:}
The goal of the key transfer stage is
to enable \TN\ to share the symmetric key $k$ with \TR\ in exchange for
collecting a signature from \TR\ that allows \TN\ to collect its reward
in later stages of the protocol. The main innovative feature of our protocol, illustrated in Figure~\ref{fig:protocol-key-transfer}, is a pair of interlocking contracts, inspired by the HTLC mechanism, that allow \TN\ to share its symmetric key ($k$) in exchange for \TR's signature on the resulting decrypted data. First, \TN\ must install contract \textit{SC-R-Sign}, which obligates \TN\ to transfer an amount \TA\ to \TR\ if the latter submits a verifiable signature over data generated by decrypting $\ELD_{\R,k}$ using $k$. Once \TR\ can see this contract installed, it must install \textit{SC-N-Key}, which promises to transfer the same amount \TA\ to \TN\ if the latter provides a valid key $k$.

Let us assume \TN\ supplies $k$ to \textit{SC-N-Key}. Decryption of $\ELD_{\R,k}$ produces $\ELD_{\R}'$, which ought to be identical to $\ELD_{\R}$ produced in the generation stage.%, but we don't know that yet).
Verification of this involves checking that the hash H$_{\ELD_{\R}}$ is indeed a hash of $\ELD_{\R}'$. If this check succeeds, it also proves that the preimage $\ELD_{\R}'$ is ciphertext that was generated by encrypting some plaintext using $pk_{\R}$; this follows from the successful verification of the zero-knowledge proof $\pi_{\ELD_{\R}}$ at the end of the Generation Stage. \TA\ is transferred to \TN\ as promised if verification succeeds; otherwise, no amount is transferred and the protocol terminates.\footnote{Stopping the protocol if $N$ produces an $\ELD_\R$ encrypted under a different public key is not mandatory. As an alternative, we can delay the validity check on $\ELD_\R$ till the final verification stage. This will make $\R$ proceed till the final step even if $\ELD_\R$ is invalid. But, the collision resistance property of the hash function will be enough to ensure honest behaviour of $\R$ and $\N$ without the need for zero-knowledge proofs.}

Now \TR\ must play its part by supplying a receipt in the form of a signature over the decrypted data $\ELD_{\R}'$ to \textit{SC-R-Sign}: $\sigma_{\R,\ELD_{\R}'}$. If it provides a valid signature that can be verified using $pk_{\R}$, amount \TA\ is transferred back from \TR\ to \TN, and the net monetary exchange is zero. If \TR\ does not supply a signature within a given time period or supplies an invalid signature, it ends up forfeiting \TA.

Last but not least, both \textit{SC-R-Sign} and \textit{SC-N-Key} contain embedded timeouts (similar but not identical to time locks). \textit{SC-N-Key} has a timeout $t$ just to ensure that \TR\ does not wait indefinitely for \TN\ to provide $k$. If $t$ expires, the protocol terminates without anyone losing money. But if \TN\ does supply a valid $k$ within $t$, \TR\ is obliged to provide its receipt signature within a timeout $t'$ encoded in \textit{SC-R-Sign}. $t'$ must be chosen to be significantly greater than $t$ so that \TR\ gets ample opportunity to supply a signature after \TN\ supplies $k$.

\begin{figure}
  \centering
  \includegraphics[width=\columnwidth]{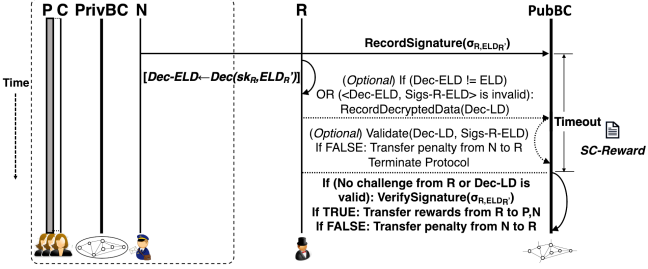}
  \caption{End-to-End Protocol: Verification and Reward Stage}
  \label{fig:protocol-verification-reward}
\end{figure}
%\makebullet{Verification and Reward Stage:}
\subsubsection{Verification and Reward Stage}
In the final stage, our protocol verifies the validity of the notification received by \TR\
and rewards \PrivBC\ members for a successful notification. To claim the reward ($\mathsf{a}$) for itself and all members of \PrivBC, \TN\ can now submit \TR's signature $\sigma_{\R,\ELD_{\R}'}$, obtained in the previous stage as evidence to \textit{SC-Reward} (see Figure~\ref{fig:protocol-verification-reward}). A timer starts upon this submission, and \TR\ is given the opportunity to challenge the veracity of the received data within a given time period (call it Timeout). \TR\ now has $\ELD_{\R}'$, which can be decrypted using $sk_{\R}$, the private key corresponding to the public $pk_{\R}$.

The produced Dec-LD ought to be well-formed, contain \LD\ and enough signatures from \PrivBC\ members to prove consensus. The set of enclosed signatures should have been created by members whose public keys were recorded to \PubBC\ in the Bootstrap Stage, and the set of signatories should match those of Sigs-\TR-ELD, supplied in the Generation Stage to prove that \TR\ was authorized to get access to \LD. Signatures in Sigs-\TR-ELD should be validated as well, given that \ELD\ (\LD\ plus signatures) has been decrypted. The value of \LD\ should be fresh and not an older version (a form of replay attack mounted by \TN).

If any of these conditions are unsatisfied, \TR\ can choose to record Dec-LD on the ledger and \textit{SC-Reward} can verify whether \TR's claims about receiving invalid data are correct. If that is the case, no reward is transferred to members of \PrivBC,
an amount is transferred from \TN\ to \TR\ as penalty for spurious information, and the protocol terminates. Otherwise, when Timeout occurs, \TR's receipt signature $\sigma_{\R,\ELD_{\R}'}$ is verified. Upon success, reward amount $\mathsf{a}$ is transferred from \TR\ to \PrivBC's members, with \TN\ getting an extra amount to cover its transaction costs in \PubBC.
%to cover for transaction costs incurred.
Upon failure, \TR\ pays no reward and \TN\ pays an amount in penalty to \TR. The protocol now terminates.

\subsection{Discussion}
We now discuss some specific aspects of \name\ in more detail.
\newline
\makebullet{Handling Multiple Concurrent Notifiers}: It is possible that
multiple \TN\ concurrently try to run our protocol for the same \LD.
To protect against this, we suggest that an additional version
attribute be attached to each \LD\ (to reject older updates to \LD)
and the smart contracts follow a first-come-first-serve policy to
select and designate an \TN.
\newline
\makebullet{Collective Authorizations}:
We encode access control rules in a smart contract \textit{SC-ACL} to ensure that a decision to allow a given \TR\ access to a given \LD passes through network consensus, hence expressing collective consent of the network members rather than a decision made by a single authority.
\newline
\makebullet{Incentive Structure}:
A central enabler for our protocol is a cryptocurrency-based public blockchain with contracts deployed on it that incentivize participants to be honest.
In \name, the intent is to ensure that \PrivBC\ members get their rewards {\it only} when they fulfill their commitments.
%To this end, crypto-currency (reward) is locked in the \textit{SC-Reward} contract during the bootstrap phase. 
%The reward can only be claimed by the members of \PrivBC\ possessing accounts in \PubBC\ when they are able to provide evidence of fulfilling their tasks. 
We lock the reward in the form of crypto-currency in the \textit{SC-Reward} contract during the bootstrap phase, which can only be claimed by members of \PrivBC\ possessing accounts in \PubBC\ when they provide evidence of fulfilling their tasks.
This is done by (i) submitting valid data in the Generation Stage, and (ii) submitting a valid signature from \TR\ in the Verification and Reward Stage. 
%Also, note that the requirement of paying transaction fees on \PubBC\ is not a disincentive for any member of \PrivBC\ to play the role of \TN\ as those costs are covered by the rewarder in the Verification and Reward Stage.
Needless to say, the reward amount should be commensurate with the perceived value of information being acquired (including any additional costs e.g. transaction fees on \PubBC) and hence left as a use case-specific configurable parameter.
\newline
%\makebullet{Cryptographic Constructs}: \makenote{Dhinakaran sprinkle this in the protocol and remove this whole thing}
%$\name$ encrypts ELD under $PK_R$ using a deterministic public-key encryption scheme to get $ELD_R$. We will use the Encrypt-with-Hash construction in Bellare et al. \cite{BBO07} with the ElGamal encryption scheme \cite{Elgamal87} as the underlying randomized public-key encryption scheme. We further encrypt $ELD_R$ using a secret key encryption scheme (AES-CTR) to obtain $ELD_{R,K}$. This second layer of encryption is to prevent $\R$ from learning $ELD_R$ (and hence LD) before the HTLC step of the protocol. This deterministic encryption provides PRIV-CCA security \cite{BBO07} and it's good enough to provide confidentiality to ELD since the underlying unforgeable signatures are `unpredictable'. 
%Here, the zero-knowledge proof $\pi_{ELDR}$, to show that $ELD_R$ is a ciphertext with $PK_R$ as the public key, is generated using a pair of discrete-log exponent checks followed by a hash check. Checking the validity of $\pi_{ELDR}$ during the Generation stage enables $R$ to stop the protocol if $N$ produces an $ELD_R$ encrypted under a different public key. As an alternative, we can delay the validity of $ELD_R$ till the final verification stage. This will make $R$ proceed till the final step even if $ELD_R$ is invalid. But the collision resistance property of the hash function will be enough to ensure honest behaviour of $\R$ and $\N$ without the zero-knowledge proofs.

\section{Analysis}\label{SEC:analysis}
%Justify the extensibility of our protocol. Suggest ways to use it for blockchain interoperability, where the recipient is not a unit but a DLT network.
%An $\name$ system has three core entities: a generator $\G$ who is a member of a private blockchian network \PrivBC{} and the receiver $\R$ who is not a member of \PrivBC{} but tries to receive notifications on a specific entry in \PrivBC{} and a notifier $\N$ who is a member of \PrivBC{} who facilitates this process.
We will now analyze $\name$ and show that it satisfies the goals listed in Section\ref{SEC:model} while relying on the assumptions listed in that section. Cryptographic primitives are expected to provide standard security properties, which we will specify as needed. Additionally, we assume that the public-key encryption scheme is robust~\cite{ABN10}. 

%We refer a notification as `valid' if there is a consensus among the members of \PrivBC{} on the message's presence in its ledger and a consensus on the receiver $\R$ being a valid recipient of the notification. These are denoted by the signatures of the members of \PrivBC{}. % and its consensus policy.
%We will classify the properties of an $\name$ system based on the entity being malicious. 

\subsection{Protecting against a malicious member of \PrivBC{}}
We now show the guarantees \name{} provides in the presence of a malicious co-owner $\C$ and/or notifier $\N$.

\subsubsection{Notification authenticity}
%\makebullet{Notification authenticity:}
$\name$ ensures that the notification received by $\R$ accurately represents $\LD$ after it has been updated in \PrivBC{}.
\begin{claim}\label{claim:NA}
A rational $R$ accepts a notification as valid only if it corresponds to a valid $\LD$ in \PrivBC{}.
\end{claim}
\emph{Proof sketch.}
The members of \PrivBC{} obtain a reward from \textit{SC-Reward} at the end of a \textit{successful} run of the protocol. This happens when $\R$ does not submit an invalid Dec-$\ELD$ to SC-Reward. At a high level, the proof of this claim will follow the following outline: the unforgeability of signatures and the soundness of zero-knowledge proofs will ensure that a rational $\R$ will not submit Dec-$\ELD$ only if it corresponds to a valid $\LD$ in \PrivBC{}.

To notify $\R$, $\N$ submits two transactions to \PubBC:
\begin{itemize}
\item $\N$ first submits $\ELD_{\R,k}$, H$_{\ELD_{R}}$, $\pi$, Sigs-$\R$-$\ELD$ to SC-Reward.
\item Then, $\N$ records a key $k$.
\end{itemize}
These actions lead to the following \textit{automatic} actions by the smart contracts which cannot be altered by $\N$ or the members of \PrivBC{} after they are installed:
\begin{itemize}
\item \textit{SC-Reward} verifies the zero-knowledge proof $\pi$ -- \textit{ensures that} $\N$ knows of an entry (i) which is an encryption of a message under the public key $pk_\R$ and (ii) whose hash is H$_{\ELD_{\R}}$
\item \textit{SC-N-Key} decrypts $\ELD_{\R,k}$ with $k$, verifies its hash with $\ELD_\R$ and finally transfers the amount \TA\ from $\R$ to $\N$ if hash is valid %(which is locked to be retrieved by \textit{SC-R-Sign}) 
-- given the above zk proof, collision resistance of $H$ ensures that $\ELD_\R$ is an encryption of some message under $pk_\R$; note that the secret key encryption scheme need not be robust \cite{ABN10}, even if $\N$ has the opportunity to provide a different key $k'$ to decrypt to a different $\ELD_\R'$ post the submission of $\ELD_{\R,k}$, since the hash collision-resistance and proof's soundness ensure that the resulting ciphertext is an encryption under $pk_\R$.
\end{itemize}
The above construct ensures that $\ELD_\R$ encrypts a message that can be decrypted using $\R$'s secret key. Here, $\R$ can decrypt and check the validity of the message, confirming that that it is a valid $\LD$ with signatures from \PrivBC{}. Even after $\R$ acknowledges receipt of $k$ from $\N$ by submitting its signature on $\ELD_\R$, as a rational actor, it can subsequently submit a decrypted $\ELD$ if it turns out to be invalid. This will ensure that $\N$ does not get the reward, and additionally is penalized.
%[If we have a penalty to N for submitting an invalid LD] Even more, a rational $\N$ will send notifications only if it expects $\R$ to accept it i.e., $\N$ takes the responsibility of verifying \PrivBC{} signatures before starting this process, else $\R$ will make $\N$ lose its amount by submitting the invalid LD.

%\paragraph{Ideal functionality} (Needs discussion) A trusted third party holds \PrivBC{}. $\G$ provides subscription list $\{(\tau_i, \R_j)\}$ to TTP. $\R$ provides approval list $\{(\tau_i, \R_j, \sig)\}$ to TTP. On an update made to $\tau$, TTP relays the update to $\R$.

\subsubsection{Information assurance}
%\makebullet{Information timeliness:}
This property of $\name$ ensures that $\R$ gets notified of any update to $\LD$.
\begin{claim}
$\name$ ensures the delivery of a notification to $R$ for each of its subscribed updates in \PrivBC{}.
\end{claim}
\emph{Proof sketch.}
This is ensured by the rationality of the members of \PrivBC{}, who are incentivized to claim a reward from $\R$ through the smart contract \textit{SC-Reward} upon sending an endorsed $\LD$. %\textcolor{red}{Especially, the extra reward promised to $\N$ incentivizes members of \PrivBC{} to act as notifiers whenever $\LD$ is updated on the ledger.}%, and (ii) predictable commit times in the \PubBC{} blockchain.

\subsection{Protecting against a malicious receiver $\R$}
We now discuss the security properties of $\name$ to protect against a malicious $\R$. 

\subsubsection{Reward fairness}
%\makebullet{Reward fairness:}
Members of \PrivBC{} and $\R$ decide on a reward to be paid by $\R$ when $\R$ receives a valid notification on $\LD$ from $\N$.  $\name$ ensures that they obtain the agreed upon reward when the notification is received by $\R$.
\begin{claim}\label{claim:RF}
Members of \PrivBC{} obtain the agreed upon reward if a rational $\R$ accepts a notification as valid.
\end{claim}
\emph{Proof sketch.}
$\R$ receives the ciphertext $\ELD_\R$ containing the notification only after $\N$ records the key $k$ on \PubBC{}. Even though $\N$ commits $\ELD_{\R,k}$, H$_{\ELD_\R}$, $\pi$ and Sigs-$\R$-$\ELD$ to \PubBC{}, $\R$ does not learn any information about $\ELD_\R$, and hence about $\ELD$, before $\N$ records $k$ due to the semantic security of the secret key encryption scheme, hiding of the commitment scheme and the zero-knowledge property of the proof system. After $\N$ records $k$, SC-N-Key transfers the amount \TA\ from $\R$ to $\N$. Since this amount \TA\ is higher than the reward that it would send to members of \PrivBC{}, a rational $\R$ would send its signature on $\ELD_\R'$ to confirm its receipt. Unless $\R$ submits decryption of an invalid Dec-$\ELD$ before the pre-decided challenge timeout, the reward is automatically transferred to the \PrivBC{} members just on $\R$'s confirmation of receipt of $\ELD_R'$. If the $\ELD_\R'$ sent by $\N$ is valid, to prevent the reward from being transferred, \TR\ should submit a Dec-$\ELD$ which upon encryption should yield $\ELD_\R'$, since we use a deterministic encryption scheme. But it is not possible for \TR\ to submit a different Dec-$\ELD$ which results in the same $\ELD_\R'$ due to the robustness of the public-key encryption scheme.

\subsubsection{Notification-receipt undeniability}
%\makebullet{Notification-receipt undeniability:}
This property prevents a malicious $\R$ from denying knowledge of $\ELD$ after it has obtained $\ELD$ through $\name$.
\begin{claim}
A malicious $\R$ will not be able to obtain knowledge of $\ELD$ while simultaneously denying receipt of the notification leading to that knowledge.
\end{claim}
\emph{Proof sketch.}
As mentioned in Claim \ref{claim:RF}, $\R$ does not learn any information about $\ELD$, and hence about LD, before $\N$ records $k$ due to the semantic security of the secret key encryption scheme and the zero-knowledge property of the proof system. Hence, $\R$ has to wait for $\N$ to record $k$ on \PubBC{}. But a rational $\R$ has to send a signature confirming its receipt of $\ELD_\R$ since we set $\mathsf{A} >> \mathsf{a}$. This signature along with the hash and zero-knowledge proof (that H$_{\ELD_\R}$ is an encryption under $pk_\R$) serve as a proof recorded on \PubBC{} that $\R$ has received a message from $\N$ that only it can decrypt.

\subsection{Fair exchange}
Now, we are ready to prove that $\name$ ensures a fair exchange between $\R$ and the members of \PrivBC{}.
\begin{claim}
$\R$ receives a valid notification if and only if the members of \PrivBC{} obtain their reward for the notification.
\end{claim}
\emph{Proof sketch.} Claim \ref{claim:RF} covers the \textit{only if} part of this claim. We will argue the \textit{if} part of the claim now. If the members of \PrivBC{} receive their reward, it means that $\R$ did not provide an invalid Dec-$\ELD$ to \PubBC{} in the Verification and Reward Stage. This implies either that $\R$ accepted the notification as valid or that \textit{SC-Reward} rejected a challenge from $\R$. The unforgeability of signatures and Claim \ref{claim:NA} ensure that a rational $\R$ received a valid notification.

\subsection{Validity of all exchanges in $\name$}
%\makenote{DB: Do we want to claim confidentiality and privacy for \PrivBC\ here? if so then reword this. else looks like this needs rewording}
%\makenote{VD: I don't think we can make such a claim here, more powerful than what's here. Claim 4.6 alone will not ensure confidentiality of data. Claim 4.6 will ensure that only exchange of a valid data to an authorized recipient will happen through our protocol. Hence, this is incentivizes some rule external to our system to be defined saying that a data is considered to have value only if it is obtained through our protocol. This will now lead to “Confidentiality of data in PrivBC” }
We now prove an interesting and powerful claim: $\name$ ensures that only authorized recipients of $\ELD$ end up possessing it. %\textcolor{red}{Only valid transactions happen through $\name$.}
\begin{claim}
%Only those $\R$ for whom PermitAccess($pk_\R$, $\LD$) had been completed successfully by \textit{SC-ACL} can receive receive valid notifications through $\name$.
Only authorized $\R$s can receive notifications through $\name$.
\end{claim}
\emph{Proof sketch.} If PermitAccess($pk_\R$, $\LD$) (in SC-ACL) had not been completed, there will not be signatures on \PrivBC{} to prove the consensus among \PrivBC{} members that $pk_\R$ is permitted to access $\LD$. The unforgeability of the signature scheme ensures that $\N$ cannot generate valid signatures on its own. Now, if $\N$ does proceed with our protocol to obtain a fair exchange of reward for the data $\ELD$, a rational $\R$ after it decrypts $\ELD_\R'$ to obtain $\ELD$ can still proceed to show the absence or the invalidity of the signatures to \PubBC{} to deny \TN\ a reward and additionally claim a penalty from it. This prevents a rational $\N$ from even initiating the transaction through our protocol if $\R$ is not permitted access to $\LD$. Additionally, the PRIV-CCA security of the deterministic encryption scheme ensures that $\ELD$ remains confidential from any observer of $\name$ messages other than members of \PrivBC\ and \TR.
%The property of $\RV$ prevents a $\R$ from receiving updates on a token which it had not been given subscription for. This property indeed assumes that the members of \PrivBC{} are honest.
%\begin{itemize}
%\item Recipient validity: for any token $\tau$, the creator of $\tau$ provides subscription for potential recipients (either as an entry in \PrivBC{}, or through external communication to $\R$). Only the permitted $\R$s can get updates on $\tau$.
%\begin{itemize}
%\item THIS PROPERTY I: This is ensured by the \PrivBC{} network attesting to the notification of $\LD$ to $\pk_R$. We assume the network ``checks'' the validity of the receiver public key before reaching a consensus on the notification for LD.
%\item No one other than a registered receiver can receive updates for a specific data token.
%\end{itemize}
%\end{itemize}

\section{Related Work}\label{SEC:related}
\textbf{Fair Exchange:}
As we stated in Section~\ref{SEC:introduction}, trusted data notification, or the exchange of a reward for authentic private network data can be modeled as a form of \textit{Fair Exchange}~\cite{asokan_fair_exchange} between the notifier and receiver, where \PubBC\ acts as the trusted third party mediating the exchange.
The exchange supported by our protocol has a few salient differences from the classic fair exchange problem that are not covered in literature: (i) one of the counterparties is a decentralized group collectively agreeing to do an exchange and allowing a single spokesperson (\TN) to execute that exchange in a way that gives both receiver and all group members their due, (ii) the item being exchanged for a reward must be kept confidential from the trusted third party, which in our protocol is a public blockchain.

\textbf{Atomic Transactions across Networks:}
HTLC~\cite{HTLC} and Atomic Swaps~\cite{herlihy2018atomic} have been proposed as techniques to swap assets between multiple parties across blockchain networks.
The InterLedger Protocol (ILP)~\cite{ILPv4} was proposed to transfer assets between entities across multiple network hops, using HTLC to transfer an asset across each hop. These mechanisms rely on counterparties having visibility into the participating network ledgers, and therefore cannot be directly applied to the trusted data notification scenario where one network is private and the counterparty (\TR) has no visibility into the that network's ledger.

\textbf{Blockchain Interoperability Frameworks:}
Relay networks like Cosmos~\cite{CosmosIBCprotocol} and PolkaDot~\cite{DOT19} enable inter-blockchain network transactions of a similar flavor to our scenario, using central blockchain as mediators.
Abebe et al.~\cite{middleware_interop} support similar interoperation, at least for data transfer, without relying on a mediating network (serving as a candidate for our \VRS\ building block).
However, while these protocols enable interoperation, none of these frameworks provide out-of-the-box support for the kind of novel trusted data notification between a private blockchain and an external party which our protocol enables.

\textbf{Cross Blockchain Data Transfer With Proofs:}
Numerous mechanism have been proposed to prove the authenticity of data shared from one blockchain to another. In public blockchains, various off-chain proof mechanisms are used to verify the validity of one public blockchain's transactions in another~\cite{SPV,zamyatinxclaim,kiayias2016PoPoW,kiayias2017NiPoWPoW}. In private blockchains, \textit{proof-of-authority}, or a quorum of signatures, are typically used to attest to the authenticity of network data~\cite{CosmosIBCprotocol,DOT19,Corda,middleware_interop}.
However, these techniques were not designed to handle trusted data notifications, and can at best act as supporting features for our novel protocol.

\section{Conclusion and Future Work}\label{SEC:conclusion}
We have presented the design of a trusted data notification protocol
for a private blockchain network to shared valid notifications
of updates to its ledger data with authorized external entities. Our solution
introduces incentives for private blockchain members to
participate in, and comply with the rules of, such a protocol.
Using a public blockchain as trusted arbitrator, standard
cryptographic mechanisms for data confidentiality and integrity,
and blockchain patterns like HTLC, our protocol guarantees
fairness for both network and recipient. We formally defined the ideal properties
of this protocol and proved that our solution satisfies those properties.
We believe our protocol is practically usable and serves as a mechanism for private
blockchain network processes to interoperate with external processes, the lack of which
has inhibited wider adoption of private blockchains in industry and government. 
In the future, we plan to build a prototype to further validate our design
and evaluate its performance for practical needs.
We will also extend the protocol to allow other business networks (built on blockchain
or other technology) to be receivers, thereby enhancing the abilities of
networks built on diverse technology stacks to interoperate with each other.

% % use section* for acknowledgment
% \ifCLASSOPTIONcompsoc
%   % The Computer Society usually uses the plural form
%   \section*{Acknowledgments}
% \else
%   % regular IEEE prefers the singular form
% \fi


%\bibliographystyle{plain}
\bibliographystyle{IEEEtran}
\bibliography{etlc_paper}

\end{document}